\documentstyle[twoside,fleqn,espcrc2]{article}

\catcode`@=11
\def\@maketitle{%                   	% Redo formatting of \maketitle
  \global\setbox\fm@box=\vbox\bgroup	% so that I can put in a preprint
    \vskip -30mm \normalsize	 	% number and date.
    \rightline{IASSNS-HEP-97/90}
    \rightline{hep-ph/9707450} 
    \rightline{July 1997}
    \normalsize
    \vskip 20mm                   
    \raggedright                  
    \hyphenpenalty\@M             
    {\Large \@title \par}         
    \vskip\@bls                   
    {\normalsize                  
     \@author \par}               
    \vskip\@bls                   
    \@address                     
  \egroup
  \twocolumn[%                    
    \unvbox\fm@box                
    \vskip\@bls                   
    \unvbox\abstract@box          
    \vskip 2pc]}                  
\catcode`@=12

\newcommand{\newc}{\newcommand}
\newc{\kkbar}{K^0-\overline{K^0}}
\newc{\eg}{{\it e.g.}}
\newc{\ie}{{\it i.e.}}
\newc{\etal}{{\it et al}}
\newc{\order}{{\cal O}}
\newc{\beq}{\begin{equation}}
\newc{\eeq}{\end{equation}}
\newc{\bea}{\begin{eqnarray}}
\newc{\eea}{\end{eqnarray}}
\newc{\gravitino}{{\widetilde{G}}}
\newc{\squark}{{\widetilde{q}}}
\newc{\mtwid}{\widetilde{m}}
\newc{\gluino}{{\widetilde{g}}}
\newc{\Qsquark}{{\widetilde{Q}}}
\newc{\usquark}{{\widetilde{u}}}
\newc{\dsquark}{{\widetilde{d}}}
\newc{\csquark}{{\widetilde{c}}}
\newc{\tsquark}{{\widetilde{t}}}
\newc{\Lslepton}{{\widetilde{L}}}
\newc{\eslepton}{{\widetilde{e}}}
\newc{\smuon}{{\widetilde{\mu}}}
\newc{\selectron}{{\widetilde{e}}}
\newc{\stau}{{\widetilde{\tau}}}
\newc{\stp}{{\widetilde{t}}}
\newc{\bino}{{\widetilde{B}}}
\newc{\half}{\frac{1}{2}}
\newc{\kev}{\,\mbox{keV}}
\newc{\gev}{\,\mbox{GeV}}
\newc{\tev}{\,\mbox{TeV}}
\newc{\mev}{\,\mbox{MeV}}
\newc{\ev}{\,\mbox{eV}}
\newc{\gsim}{\lower.7ex\hbox{$\;\stackrel{\textstyle>}{\sim}\;$}}
\newc{\lsim}{\lower.7ex\hbox{$\;\stackrel{\textstyle<}{\sim}\;$}}
\newc{\mz}{m_Z}
\newc{\mpl}{M_{Pl}}
\newc{\mmess}{M}
\newc{\vev}[1]{\left\langle #1\right\rangle}
\newc{\misset}{/\!\!\!\!E_T}
\renewcommand{\bar}{\overline}
\def\NPB#1#2#3{Nucl.\ Phys.\ {\bf B#1}, #2 (#3)}
\def\PLB#1#2#3{Phys.\ Lett.\ {\bf B#1}, #2 (#3)}
\def\PLBold#1#2#3{Phys.\ Lett.\ {\bf#1B}, #2 (#3)}
\def\PRD#1#2#3{Phys.\ Rev.\ {\bf D#1}, #2 (#3)}
\def\PRL#1#2#3{Phys.\ Rev.\ Lett.\ {\bf#1}, #2 (#3)}
\def\PRT#1#2#3{Phys.\ Rep.\ {\bf#1}, #2 (#3)}

\begin{document}

\title{Gauge-Mediated Supersymmetry Breaking: \\[-3pt]
	{\large Introduction, Review and Update}}

\author{Christopher Kolda\address{School of Natural Sciences,
        Institute for Advanced Study, Princeton, NJ 08540, USA}
        \thanks{This work was partially supported by DOE grant 
	\#DE-FG02-90ER40542.}}

\begin{abstract}
Recent progress in the gauge-mediated supersymmetry breaking is
reviewed, with emphasis on the theoretical problems which
gauge-mediated models so successfully solve, as well as those problems
which are endemic to the models themselves and still beguile theorists today.
(Talk given at the $5^{\rm th}$ International Conference on Supersymmetries
in Physics (SUSY-97), Philadelphia, PA, May 27-31, 1997.)
\end{abstract}

\maketitle

For all the theoretical successes of supersymmetry (SUSY), its one
overriding problem is, and has been, that there is no direct experimental
evidence for it. But rather than throw SUSY out,
we can try to retain some of its desirable properties (\eg, solution
to the hierarchy/naturalness problems, unification of the gauge couplings)
by building theories with {\sl spontaneously broken}
(or ``softly-broken'') $N=1$ SUSY.

The central problem of SUSY would then appear to be ``How is SUSY
broken?'' For many years this was the prime focus of research into
SUSY. Two realizations have pushed that question somewhat to the sidelines.
First new exact results in SUSY
have led to large numbers of new models in which SUSY
is spontaneously broken. In the past three years alone, the number of such
models has increased exponentially~\cite{talk:nelson}. Now it appears that
spontaneous SUSY-breaking is not a peculiar behavior exhibited by only a few
special models, but is in fact rather generic.

Second, it was realized many years ago~\cite{dg}
that if the mechanism for SUSY-breaking coupled too closely to the
Standard Model (SM) spectrum, that spectrum would have to exhibit certain
sum rules, namely
$\mbox{STr}\,M^2=0$,
which holds for sets of states with identical conserved quantum numbers.
This relation implies that one of the $u$-type squarks must be lighter 
than the $u$-quark, in clear contradiction to experiment.
However there is a way out of this phenomenological disaster, for
this equation holds only at tree level, and only for renormalizable
theories. The possibility then exists to break SUSY dynamically, but in some
sector which only couples to the SM sector via loops or via
non-renormalizable operators. We then refer to the SUSY-breaking sector
as ``hidden,'' the SM sector as ``visible,'' and the intermediate states,
which appear in loops or are integrated out to produce the
non-renormalizable interactions, as ``messengers'' or ``mediators.''

The banishment of SUSY-breaking to some mysterious ``hidden sector'' has
an immediate phenomenological consequence. It is no longer the mechanism
of spontaneous SUSY-breaking itself that prescribes the form which 
soft SUSY-breaking will take in the visible sector, but rather it is the
mechanism for mediating SUSY-breaking that ultimately determines
how SUSY will manifest itself in our colliders.
The central problem of SUSY has changed. It is now ``How does the Standard
Model find out about SUSY-breaking?''

\section{Communicating SUSY-breaking}

There have evolved several approaches to dealing with the question of 
how SUSY-breaking is communicated to the SM. The first~\cite{dadt}
is simply to admit ignorance and to parametrize that ignorance with
the most general effective Lagrangian 
consistent with the symmetries of the SM and with softly-broken
$N=1$ SUSY. A special case of this most general Lagrangian is the one
which is minimal, that is
the one which requires the least extension of the SM
in order to accomodate the broken SUSY. This is the Minimal Supersymmetric
Standard Model (MSSM). I call this approach ``Don't ask, don't tell.'' The
``don't ask" part is obvious, for we simply don't worry about
the exact nature of how SUSY-breaking is mediated. The ``don't tell'' is
the unfortunate part, for even in the MSSM (emphasis on ``minimal'')
there are 106 unknown parameters~\cite{ds}:
26 masses, 37 angles and 43 phases.
With so many unknowns, extracting unequivocable experimental predictions
for SUSY has proven to be a risky undertaking~\cite{risky}.

But already in this very general framework, we can begin to see that some
organizing principle must be in place. With arbitrary, random parameters
the MSSM has serious problems with flavor (and also CP) violation. Put simply,
once SUSY is broken, sparticles and particles need not be diagonal in the
same eigenbasis. That is, the rotations which take the fermions from their
interaction to their mass eigenstates need not do the same for their
superpartners. If we denote these unitary rotation matrices as $U$ and
$\widetilde{U}$ for the fermions and scalars respectively, then
the interation of a 
gaugino with a fermion and its scalar partner picks up a factor $U^\dagger
\widetilde{U}\neq{\bf1}$, leading to flavor-changing neutral currents (FCNCs).

These FCNCs show up in a variety of processes, though always at loop
order: $\kkbar$ mixing, $\mu\to e\gamma$, etc. Consider the
first of these~\cite{kkbar}. 
%In Fig.~\ref{fig:kkbar}\ I show the leading contribution to $\kkbar$ mixing, 
%where I am working in the gluino interaction eigenbasis. 
Requiring that the SUSY contributions to $\Delta m_K$ be smaller
then the its experimental value yields the well-known contraint:
\beq
\mbox{Re}\,\left\lbrace\frac{{\cal A}^2}{\mtwid^2}\,\left(\frac{\delta
m_\Qsquark^2}{\mtwid^2}\right)^2\right\rbrace\lsim5\times10^{-9}\gev^{-2}
\eeq
where $\mtwid$ is some average squark mass, $\delta m^2_\Qsquark$ is
the $\widetilde{d_L}-\widetilde{s_L}$ mass difference, and ${\cal A}$
is a function of the angles which rotate $U\to\widetilde{U}$. For
generic $U$ and $\widetilde{U}$ the average value for ${\cal A}^2\simeq
1/20$.

The constraints on CP violation in the kaon system
%coming from the imaginary piece of Fig.~\ref{fig:kkbar}
are even stronger. Demanding
that the SUSY contributions to $\epsilon_K$ not exceed its experimental
value yields:
\beq
\delta\,\left|\frac{{\cal A}^2}{\mtwid^2}\,\left(\frac{\delta
m_\Qsquark^2}{\mtwid^2}\right)\left(\frac{\delta
m_\dsquark^2}{\mtwid^2}\right)\right|\lsim10^{-13}\gev^{-2}
\label{eq:cp}
\eeq
where $\delta$ is the appropriate squark phase. Using Eq.~(\ref{eq:cp})
with $\delta\sim\order(1)$, ${\cal A}^2\simeq1/20$ and $\mtwid=500\gev$,
one finds that $m_{\widetilde{s}}-m_{\widetilde{d}}\lsim200\mev$. Or
for $\delta\mtwid^2/\mtwid^2\sim\order(1)$, we
must have $\mtwid\gsim700\tev$! Thus FCNC and CP constraints demand
either strong mass degeneracies among the scalars,
or scalars so heavy that they decouple from the relevant processes
altogether.

(There are at this time several competing proposals for solving this
flavor/CP problems, including enforced mass degeneracies, decoupling of
sparticles, and ``alignment'' in which $U=\widetilde{U}$ and so
${\cal A}=0$. I will only be discussing the first of these three
possibilities in this talk.)

Within a most general Lagrangian, one has no way to approach the FCNC/CP
problems; one has to consider specific mechanisms
by which SUSY-breaking may be communicated from the hidden to the visible
sector. The canonical method is supergravity.
Local SUSY, {\sl a.k.a.} supergravity,
mixes the hidden and visible sectors through gravitational
interactions (the mixing goes to zero as $\mpl\to\infty$). The
mixing terms, being non-renormalizable, suppress the scale of SUSY-breaking
in the visible sector from that in the hidden sector. For example, for
an O'Raifeartaigh-type breaking in the hidden sector at some scale
$\sqrt{F}$, the apparent scale of SUSY-breaking in the visible sector
will be $F/\mpl\ll\sqrt{F}$.

The story of supergravity (SUGRA) is well-known. Mass universalities seem
to naturally fall out without inputting them, solving both the FCNC
and CP problems~\footnote{Mass universalities solve the SUSY CP problem
in the kaon system; however, they do not solve another SUSY CP problem
which strikes in various electric dipole moments. This second CP problem
will be discussed in Section~\ref{sec:open}.}. 
One is left with only 5 free parameters
at the SUGRA scale: a common scalar mass $m_0$, gaugino mass
$M_{1/2}$, $A$-term $A_0$, $B$-term $B_0$, and of course $\mu$.

The truism that usually accompanies these universalities is that ``gravity
is flavor-blind.'' I agree. Gravity has no reason to arrange its interactions
so that they are diagonal in the same basis in which the Higgs couples to the
fermions --- and for that very reason it is clear that any basis chosen
by gravity will differ from the Higgs interaction basis, leading to
mass non-universalities and FCNCs. Even if gravity chooses no preferred
basis at tree level, non-renormalizable terms in the K\"ahler potential,
corrections coming from string interactions, renormalization group flow, and
any of a dozen other possible effects will choose a preferred basis for
the communication of SUSY-breaking, leading to disaster. In particular, if
there is any source of flavor physics (\eg, any interactions which
differentiate between $d$- and $s$-quarks) between the Planck scale and
the weak scale, the mass universalities will be destroyed.

It would seem preferable for the mass degeneracies among the squarks and
sleptons, rather than be accidents, to be guaranteed by the nature of the
mediation mechanism. In particular, one would like the communication
mechanism to respect global symmetries under which the various $u$-type
quarks are identical, likewise for the various $d$-type quarks and the 
various leptons.

We don't have to go far to find such symmetries, for they are
the gauge symmetries of the SM. If the scalar soft masses were
functions only of the gauge charges of the individual sparticles,
universality would be automatic.  And if the scale at which the communication
of SUSY-breaking takes place is well below the Planck scale, then those
Planckian ``corrections'' which upset universality in SUGRA cannot
disrupt it here.

\section{Gauge Mediation}

The defining principle of gauge mediation, and the reason it solves
the flavor problem, is that the SUSY partners of SM fermions receive the 
dominant part of their masses via gauge interactions. 

The canonical model for gauge mediation was developed in the early 
1980's~\cite{early}. Suppose there exist some set of states (superfields)
$\hat\psi$ with SM couplings but which are not 
part of the MSSM spectrum.
Since the fermionic components of the $\hat\psi$ superfields must be heavy
and not contribute to the SM anomalies, take them to be vectorlike with
respect to the SM gauge interaction (\ie, $\hat{\bar\psi}\hat\psi$ is an
SM singlet). In the part of the superpotential
responsible for mediation, couple the $\hat\psi$ 
fields to a new singlet superfield $\hat X$:
\beq
W_M=\lambda \hat X\hat{\bar\psi}\hat\psi+\cdots
\label{eq:wm}
\eeq
where $\hat X$ receives both $A$ and $F$ vevs. (Alternatively, the $A$ vev
of $\hat X$ can be replaced with an explicit mass.) The source of
the $F$ vev in the hidden sector is essentially irrelevant for most of the 
phenomenology and we take it as simply given; the exception is in the
couplings of the gravitino to be discussed in Section~\ref{sec:pheno}.

$W_M$ of Eq.~(\ref{eq:wm}) induces masses for the fermionic components of
$\hat\psi$ (which I denote $\widetilde\psi$) 
setting the overall scale for the messenger sector:
\beq
\mmess\equiv m_{\widetilde\psi}=\lambda X
\eeq
while the scalar mass matrix has the form:
\beq
m^2_{\psi,\bar\psi}=\left(\begin{array}{cc} \lambda^2 X^2 & \lambda F_X \\
\lambda F_X & \lambda^2 X^2 \end{array}\right).
\label{eq:matrix}
\eeq
The eigenvalues of Eq.~(\ref{eq:matrix}) are:
\beq
m^2_{\psi,\bar\psi}=\mmess^2\pm\lambda F_X = \mmess^2(1\pm x)
\eeq
where $x\equiv\lambda F_X/\mmess^2$; $x<1$ 
in order to have positive squared masses.
We define one more scale:
\beq
\Lambda\equiv\frac{\lambda F_X}{\mmess}
\eeq
which, as we will see, controls the weak scale.

Because the $\hat\psi$ fields are charged under the SM gauge groups, the
gauginos of the MSSM can receive masses through loops of these new fields.
In particular, the mass matrix of Eq.~(\ref{eq:matrix}) contributes to the
gaugino masses at 1-loop. Gaugino
$\lambda_i$ of SM group $G_i$ receives a mass:
\beq
M_{\lambda_i}(\mmess)=\frac{\alpha_i}{4\pi}\,g(x)T_i(\psi)\Lambda
\label{eq:gaugino}
\eeq
where
$T_i(\psi)$ is the Dynkin index of the representation of $\psi$ under
$G_i$ and $g(x)=1+x^2/6+\order(x^4)$.

The scalars of the MSSM do not receive soft masses until 2-loop order.
For scalar $\phi$:
\beq
m^2_{\phi}(\mmess)=2f(x)\Lambda^2\sum_i\left(\frac{\alpha_i}{4\pi}\right)^2
C_i(\phi)T_i(\psi)
\label{eq:scalar}
\eeq
where $C_i(\phi)$ is the quadratic Casimir of the representation of $\phi$ 
under $G_i$ and $f(x)=1+x^2/36+\order(x^4)$. Recall that there are also
$D$-term contributions to scalar masses not included here.
[An aside on conventions:
I am using an $SU(5)$ normalization for $\alpha_1=\frac{5}{3}\alpha_Y$
and so $T_1(\varphi)=C_1(\varphi)=\frac{3}{5}Y^2$ for any field $\varphi$.
For non-abelian $G_i=SU(n)$, $T_i(\varphi)=\frac{1}{2}$ and 
$C_i(\varphi)=\frac{n^2-1}
{2n}$ for $\varphi$ in the fundamental representation.] For the 
oft-cited case in which $(\psi,\bar\psi)$ are $N$ pairs of 
$({\bf5},{\bf\bar5})$ of $SU(5)$ and $x\ll1$, we
notice that gaugino masses scale as $N$, while scalar masses scale as
$\sqrt{N}$.

Finally, the trilinear soft terms ($A$-terms) arise at 2-loops. Since they
have mass dimension-1, they are small compared to the rest of the soft
masses and thus one can take $A(\mmess)\simeq0$.
The case of the bilinear, dimension-2 $B_\mu$ term will be discussed later.

One success of this particular
mechanism for gauge mediation is that while
gaugino masses arise at 1-loop, scalar mass-{\sl squareds} arise at 2-loops.
Thus gaugino and scalar masses are the same order in $\alpha$. This is 
a noteworthy, because the simplest models of gauge
mediation (those without the messenger fields) typically give
masses to scalars at lower order than to gauginos, producing models
with ultra-light gluinos.

One should make note of the scales that play a role in gauge mediation.
Because LEP constrains the selectron mass $m_\eslepton>45\gev$, then
$\Lambda>30\tev$. If by some fine-tuning argument we demand for gluinos
that $M_3\lsim1\tev$, then $\Lambda\lsim120\tev$. (These assume
one pair of ${\bf5}+{\bf\bar5}$ messenger fields.) If all couplings in the
problem are $\order(1)$ then all mass scales will be $\order(\Lambda)$, 
far below the Planck scale. Thus there will be no problems
induced by supergravity corrections. (In more general models, the
scales can differ greatly from one another. Then the requirement that
supergravity corrections be small translates into the bounds
$F_X\ll\mz\mpl$ and $\mmess\lsim10^{15}\gev$.)

\section{Dine-Nelson Models} \label{sec:dn}

In the last few years, attention has been drawn back to gauge mediation as a
viable alternative to supergravity mediation. Much of that renewed interest
has been sparked by a series of models proposed by Dine, Nelson, and
collaborators~\cite{dnns}. Because these models demonstrate both the successes
and failings typical of models of gauge mediation, I will highlight their
structure briefly.

Dine-Nelson models are divided into a tower of sectors, beginning with the 
hidden sector in which SUSY is broken spontaneously by some strong
gauge dynamics. One often locates some non-anomalous
global symmetry of the hidden sector which can be weakly gauged.
That ``messenger group,'' typically a $U(1)$, then communicates
SUSY-breaking to a set of fields, $\varphi^\pm$, which are SM singlets, giving
them negative squared masses. The $\varphi^\pm$ then couple to the messenger
singlet $X$ through terms in the superpotential:
\beq
W_M=k\hat X\hat\varphi^+\hat\varphi^- + \lambda \hat X\hat{\bar\psi}\hat\psi
+k'\hat X^3.
\eeq
Setting $\lambda=0$ for now, the potential $V(\varphi^\pm,X)$ is minimized
when $\vev{\varphi^+\varphi^-}$, $\vev{X}$ and $\vev{F_X}$ are all non-zero.
Keeping that solution once $\lambda$ is turned back on, the mass matrix of
Eq.~(\ref{eq:matrix}) is reproduced and SUSY-breaking is communicated to
the MSSM through gauge interactions.

What are the successes of the gauge mediation approach? 
First, FCNCs are absent because of the scalar mass degeneracy. 
Second, scalar and gaugino masses are roughly the same size. 
Third, electroweak symmetry breaking occurs
radiatively just as it does in supergravity models. And last, the models are
highly predictive, with the entire spectrum of soft masses 
determined (to a good approximation) from just one input: $\Lambda$.

%The third success above requires some comments. In supergravity models
%it is well known that the $H_U$ Higgs field, though having positive
%mass-squared at the Planck scale, can acquire negative mass-squared at the 
%weak scale through loops of top (s)quarks:
%\beq
%\frac{dm^2_{H_U}}{d\log Q}\simeq \frac{1}{8\pi^2}\left\lbrace -3g_2^2M_2^2
%+6y_t^2m_\stp^2\right\rbrace.
%\eeq
%Since $y_t>g_2$, $m^2_{H_U}$ is driven negative in the 
%infrared. However, $m^2_{H_U}\simeq m^2_\stp$ so it happens slowly, requiring
%the 30 e-foldings present in supergravity models 
%to get finally push $m^2_{H_U}$ below zero. In 
%gauge-mediated models where $\mmess\ll\mpl$, there are very few
%e-foldings in which to push $m^2_{H_U}$ negative. Radiative electroweak
%symmetry breaking is rescued by the fact that in gauge mediation
%$m^2_{H_U}\ll m^2_\stp$, driving $m^2_{H_U}$ negative in only a few
%e-foldings.

\section{Open Theoretical Issues} \label{sec:open}

Despite their successes, the
specific models of gauge mediation in the last section are open to a 
number of possible criticisms.

~

\noindent{\it The global minimum of the scalar
potential breaks color and not SUSY.}
The minimization performed in the last Section of the messenger
potential $V(\varphi^\pm,X)$ was not really
correct for $\lambda\neq0$. With non-zero $\lambda$, the minimum
of the potential $V(\varphi^\pm,X,\psi,\bar\psi)$ occurs at
$\vev{\bar\psi\psi}=-\frac{k}{\lambda}\vev{\varphi^+\varphi^-}$ and
$\vev{F_X}=\vev{X}=0$. That is, $\psi$ does not find out about SUSY-breaking,
but instead gets a non-zero vev, breaking $SU(3)\times U(1)$.
The minimum in which SUSY is broken is only local, not global~\cite{ddr}.

There have been a number of suggestions for circumventing 
this problem. There are ``cosmological'' solutions to the problem, \ie, perhaps
the present universe exists in a long-lived, metastable vacuum in which
QCD and QED are preserved but SUSY is broken. There are also particle
physics solutions. For example, 
we could add explicit masses for the $\psi$ messenger fields
to push their vevs to the origin~\cite{randall}. 
Such masses may seem {\it ad hoc} but
as long as they don't have to sit at any one special scale, they are not 
unnatural. (In fact, they can sit comfortably at any scale between $100\tev$
and $10^{15}\gev$.) If the messenger group is a $U(1)$,
it is also possible to add extra matter which is chiral
with respect to it~\cite{ddr}. 
Such extra matter can push the position of the minimum back to where we 
want it. However such matter can induce Fayet-Iliopoulos terms in the
hidden sector, in which case it is known that there is no way to protect
the messenger $U(1)$ from mixing with hypercharge and destabilizing the
visible sector~\cite{dkm}\ (see below).

~

\noindent{\it Messenger U(1) interactions can be dangerous.}
For $U(1)$ gauge groups, the field strength $F_{\mu\nu}$ is gauge-invariant.
Thus for theories with two (or more) $U(1)$ groups, the gauge kinetic
pieces of the Lagrangian can mix: ${\cal L}\sim F_{a\mu\nu}F_b^{\mu\nu}$ for
$U(1)_a\times U(1)_b$. Because such terms are renormalizable, they can
be induced by Planck-scale physics. Unless there is a symmetry to prevent
such terms, one should in fact expect them to be present.
And in SUSY, if the gauge kinetic pieces mix, then the $D$-terms must also.

Identify hypercharge as one of the $U(1)$ factors and assume the other
is in the hidden sector, as in the model of Section~\ref{sec:dn}. If 
{\it (1)} kinetic mixing occurs between the two $U(1)$'s, and {\it (2)} the 
$D$-vevs of the hidden sector $U(1)$ are of order the scale of SUSY-breaking
in that sector, then the large hidden sector $D$-term will be 
communicated to the scalars of the MSSM, pulling their masses up to the
scale of hidden sector SUSY-breaking~\cite{dkm}.

There are two ways to get around this problem. The first is to find a
charge-conjugation symmetry which acts on only one of the $U(1)$'s, namely
$C:A_a^\mu\to-A_a^\mu$ while $C:A_b^\mu\to A_b^\mu$. Such a symmetry will
forbid kinetic mixing, particularly if it is a gauged discrete symmetry.
The second possibility is to work
only with non-abelian messenger groups for which kinetic mixing cannot occur.

~

\noindent{\it The $\mu$-problem is worse than usual.}
The $\mu$-problem of the MSSM is familiar. If $W=\mu H_UH_D$ is a
SUSY- and $G_{SM}$-invariant mass term, why is $\mu\sim
\mz$ instead of $\mu\sim\mpl$? We know that 
$\mu\sim\mz$ because it contributes to the Higgs potential and thus
to the $Z^0$ mass:
\beq
\mu^2=\frac{m^2_{H_D}-m^2_{H_U}\tan^2\beta}{\tan^2\beta-1}-\frac{1}{2}
\mz^2
\label{eq:mu}
\eeq
where all the masses on the RHS are $\order(\mz)$. Within the context
of supergravity, the most promising solution is the
Giudice-Masiero mechanism~\cite{gm}\ in which the hidden and visible
sectors mix through a non-minimal K\"ahler potential,
$K=K_0+\frac{1}{\mpl}X^\dagger H_UH_D+h.c.$, where $K_0$ is the canonical
piece and $X$ is some hidden sector field with
$F_X\sim\mz\mpl$. Then:
\beq
\int d^4\theta\, \frac{\hat X^\dagger}{\mpl}\hat H_U\hat H_D
=\int d^2\theta\, \frac{F_X}{\mpl}\hat H_U\hat H_D,
\eeq
which contributes to the superpotential with coefficient
$F_X/\mpl\sim\mz$ and is the usual $\mu$-term. In gauge-mediated models
this mechanism cannot work for the same reason that other supergravity
contributions are small: $F_X/\mpl\ll\mz$.

Gauge-mediated models also 
have a second, related problem which does not show up
in supergravity. The dimension-2, bilinear soft term $B_\mu H_UH_D$ does
not arise until 3-loops or beyond. It is therefore, like the $A$-term, 
essentially zero. This is not phenomenologically feasible since the
pseudoscalar Higgs of the MSSM gets a mass
$m_A^2=2B_\mu/\sin2\beta$.
As $B_\mu\to0$, $m_A\to0$ and $A^0$ becomes an axion. 

The oldest solution to the $\mu$-problem is to extend the MSSM to the NMSSN,
which is the MSSM plus a singlet with superpotential
\beq
W=\lambda_H \hat S\hat H_U\hat H_D + \kappa \hat S^3
\label{eq:nmssm}
\eeq
Then $\vev{S}$ provides the $\mu$-term. An obvious candidate for that singlet 
is the field $X$ which appears in $W_M$~\cite{dnns,dgp}. Here $\vev{X}$ would
provide a $\mu$-term and $\vev{F_X}$ would provide a $B_\mu$ term, seemingly
as desired. But since $\vev{X}\gg\mz$, a solution
to the $\mu$-problem clearly requires $\lambda_H\ll1$. Then
\bea
\mu&=&\lambda_H X \\
B_\mu&=&\lambda_H X^2\gg\lambda_H^2 X^2
\eea
Thus $B_\mu\gg\mu^2$. If we choose $\mu\sim\mz$ then $B_\mu$ is huge, causing
the Higgs potential to become unbounded from below. If we choose $B_\mu
\sim\mz$ then $\mu$ is tiny, leading to light charginos that would have been
found at LEP.

One could try to avoid the problems intrinsic to $X$ by introducing a new
singlet $S$ just to solve the $\mu$-problem~\cite{dnns,dgp}. 
Unfortunately, being a gauge
singlet, $S$ does not receive a very large soft mass and so its physical
component is very light. Another way to
see this is to notice that the superpotential of Eq.~(\ref{eq:nmssm})
has an $R$-symmetry which is only broken at 2-loops by small $A$-terms.
Thus the light field is really
an $R$-axion. 

Finally, it has been suggested that non-renormalizable interactions in the
superpotential could conspire to produce $\mu\sim\mz$~\cite{dnns}. Such
models again yield $B_\mu\ll\mz^2$.
%For example, the
%K\"ahler potential and superpotential~\cite{dnns}:
%\bea
%K&=&K_0+\frac{1}{\mpl^2}\hat X^\dagger \hat X\hat S^\dagger \hat S \\
%W&=&\frac{1}{\mpl^2}(\hat X\hat S^4+\hat S^5)+\frac{1}{\mpl}\hat S^2
%\hat H_U\hat H_D
%\eea
%yield $\mu\sim\mz$ but again $B_\mu\ll\mz^2$.

~

\noindent{\it There is still a CP/electric dipole problem.}
Despite the fact that the CP problem in the kaon system has been resolved 
by the mass degeneracies, there is another CP problem which persists.
by degeneracies. In the simplest models of gauge mediation, there
remain 4 non-zero CP-violating phases beyond those of the SM: 
$\arg(\mu)$, $\arg(A)$, $\arg(B_\mu)$ and $\arg(M_3)$. Of these
only two combinations are physical~\cite{cpv}:
\beq
\Phi_A=\arg(A^*M_3)\mbox{~~and~~}
\Phi_B=\arg(B^*_\mu\mu M_3).
\eeq
These phases contribute to electric dipole moments of quarks and leptons,
and in turn, nuclei. One finds for the neutron electric dipole moment:
\beq
d_N\simeq 2\left(\frac{100\gev}{\mtwid}\right)^2\sin\Phi_{A,B}\times
10^{-23}\, e\,\mbox{cm}
\eeq
for some generic squark mass $\mtwid$. Experimentally
$d_N<1.1\times 10^{-25}\, e\,$cm. Thus $\order(1)$ phases are only allowed
if $\mtwid\gsim1\tev$. Any solution which doesn't include very heavy
squarks must compensate by finding some way to enforce small phases.

\section{The Minimal Messenger Model}

Recall now the earlier discussion of the $\mu$- and $B_\mu$ problems.
It should be clear that in gauge-mediated models, the solutions to these
two problems are intimately connected. And unfortunately we usually can
solve one only at the expense of the other. Now we will show that certain
types of gauge-mediated models actually solve both problems simultaneously,
while at the same time resolving the CP/electric dipole problem and
being highly predictive.

Start from a gauge-mediated model in which the $\mu$-problem has
been solved at the expense of small $B_\mu\ll\mu^2$. As was
said before, finite (threshold) contributions to $B_\mu$ only come in
at 3-loops and thus are very small. However the 
divergent contributions receive log enhancement. In supergravity,
such an enhancement is huge; here it is relatively small, but
enough to lift the axion to experimentally allowed masses~\cite{bkw}.

There are two symmetries of the MSSM which try
to enforce $B_\mu=0$: a Peccei-Quinn symmetry, broken by non-zero $\mu$,
and an $R$-symmetry, broken by gaugino masses. Thus we expect contributions
to $B_\mu$ proportional to $\mu M_{\lambda_i}$. This
in fact happens through the 1-loop RGE's for $B_\mu$ which have an
approximate solution (for $\mmess=100\tev$): 
\bea
B_\mu(\mz)&\simeq&\mu M_2(\mmess)\left[-0.12+0.17y_t^2\right] \\
&\sim&\alpha_2\mu M_2.
\eea
For $\mu\sim M_2\sim\mz$ we can write $\tan\beta\sim\mz^2/B_\mu\sim
\alpha_2^{-1}\simeq30$.
The ``axion'' mass is $m_A\simeq400\gev$. (More precise calculations
including the full effective potential and threshold effects yields~\cite{mmm}
$\tan\beta\simeq50$.) This highly predictive 
version of gauge-mediation, in which $B_\mu$ is
set to zero at the messenger scale and is therefore no longer a free
parameter, has been called the Minimal Messenger Model (MMM)~\cite{bkw,mmm}.

Such large values for $\tan\beta$ will strike those familiar with
supergravity as unnatural, following the arguments of Ref.~\cite{nr}.
However those arguments work only for models with one mass scale. 
Such is not the case here, where the mass scales of
interest are related but very different: $B_\mu\sim\alpha_2
\mu M_2$. The hierarchy in the vevs is then simply a reflection in the
hierarchy of the mass scales. 
%This is, to my knowledge, the only
%SUSY model in existence with naturally large $\tan\beta$. 

The MMM has one other great advantage. Because $B_\mu(\mmess)=A(\mmess)=
0$, then too $\Phi_A(\mmess)=\Phi_B(\mmess)=0$. While the first relation
is modified by RGE's when running down to the weak scale, the second is
not. This is because:
\beq
\frac{dB_\mu^*}{d\log Q}\propto M_i^* \quad\mbox{and}\quad 
\frac{dA^*}{d\log Q}\propto M_i^*
\eeq
which renders $\Phi_{A,B}=0$ a stable fixed-point. 
Thus the CP/electric dipole
problem has been solved by finding a natural way in which to suppress
the new SUSY phases without having to enforce very heavy sparticle
spectra.

\section{Direct Transmission Models}

In trying to overcome some of the difficulties of the Dine-Nelson models
highlighted in Section~\ref{sec:dn}, some natural questions arise, such
as ``Can we live without a messenger interaction?'' and ``Can $\psi$
live in the SUSY-breaking sector itself?'' The answer to both questions
is ``yes'' provided the SUSY-breaking sector possesses
an anomaly-free global symmetry large enough to embed the entire SM gauge
structure and which is not broken concurrent to SUSY breaking. Such
models are called ``direct transmission'' models.

Finding such large global symmetries in the hidden sector seems
to be quite a difficulty, for it generally requires very large
dynamical SUSY-breaking groups such as $SU(15)$ or $SU(7)\times SU(6)$.
The large size of the dynamical groups feeds back into the visible sector
because we want the $\psi$ to transform under the fundamental
representation of the dynamical groups. Thus a model whose dynamical
group is $SU(n)$ containing a global $SU(5)$ will have $n$ copies of
${\bf5}+{\bf\bar5}$. If $\sqrt{F}\sim\mmess\sim50\tev$, then perturbativity 
up to the Planck scale requires $n<4$, which is too small.

The solution to this conundrum is obvious: move the messenger mass scale
far abovet the weak scale, and consequently, far above the SUSY-breaking
scale $\sqrt{F}$. The means to get this is less obvious~\cite{ptamm}. 
In most Dine-Nelson
type models, the scalar potential of the messenger sector has no flat
directions, and so on minimization one finds $F_X\sim X^2$ for any field
$X$ participating in SUSY breaking. However, along a flat direction the
potential doesn't turn up until very large values of $X$ so that at the 
minimum $F_X\ll X^2$. This is precisely the desired behavior. One caveat
however: we need to ensure that the supergravity contributions remain small 
compared to the gauge-mediated contributions. Thus we must have:
\bea
\vev{X}&\lsim&10^{15}\gev \\
\frac{F_X}{\mpl}&\lsim&\sqrt{n}\,\frac{\alpha}{4\pi}\,\frac{F_X}{X}
\eea
for dynamical $SU(n)$ with gauge coupling $\alpha$.

%To date, direct transmission models have been built using the following
%dynamical$\times\{\mbox{flavor}\}$ groups: 
%$SU(13)\times SU(11)\times \{SP(10)\supset
%SU(5)\}$~\cite{pt}, $SU(7)\times SU(6)\times \{SU(5)\}$~\cite{amm}, and
%$SP(8)\times\{SU(10)\supset SU(5)\}$~\cite{murayama}.

Direct transmission models have most, if not all, of the following
properties: {\sl (i)} Gauge coupling unification occurs just as in the MSSM.
In particular, there is no loss of perturbativity before the unification
scale. {\sl (ii)} There are no gauge singlets in the model.
{\sl (iii)} The SUSY-breaking, QCD/QED-preserving minimum
is global. {\sl (iv)} The squarks are no longer quite so heavy compared to
the sleptons. The large amount of running to go
from $Q=\mmess$ to $\mz$ washes out the largest mass hierarchies among
the sparticles. {\sl (v)} The gravitino has a mass $m_{3/2}\sim$ few GeV.
This may be a serious cosmological problem (see Section~\ref{sec:cosmo}).
{\sl (vi)} The detector signatures will very closely resemble those of
supergravity models. In particular, the NLSP decay to gravitinos occurs
outside the detector (see Section~\ref{sec:pheno}). 
Finally, {\sl (vii)} $\mbox{STr}\,\mmess^2>0$
where the supertrace is only over the set of $\psi+\bar\psi$ messenger fields.

The impact of that last statement is only appreciated when one considers the
2-loop running of the MSSM scalar masses. There one finds a correction
to the soft scalar masses of the form~\cite{ptamm}:
\beq
\delta m^2_\phi=-\frac{\alpha^2_i}{4\pi^2}C_i(\phi)T_i(\psi)\,
\mbox{STr}\,\mmess^2\,\log\frac{\Lambda_{UV}}{m_{\widetilde{\psi}}}.
\eeq
In the direct transmission
models, the messenger sector contains a number of $\psi$-fields and there may
be a sizeable hierarchy which develops among them. Then $\Lambda_{UV}$ is
to be interpreted at the largest mass in the messenger sector. If
$\mbox{STr}\,\mmess^2>0$ and some $m_{\widetilde\psi}\ll\Lambda_{UV}$
then $\delta m^2_\phi$ will push MSSM scalars to negative squared
masses. Such is the case in the models of Refs.~\cite{ptamm}, but not
that of Ref.~\cite{murayama}. For all three models, the bulk of the
messenger sector sits near $10^{15}\gev$ while a few fields sit near
$100\tev$; in the last case, however, those light fields are eaten as
the flavor symmetry breaks down from $SU(10)$ to $SU(5)$.

%As of this conference, the direct transmission models are very new. One can
%expect that many of their features will be more fully explored and 
%explained next year at SUSY-98.

\section{Phenomenology and the Gravitino} \label{sec:pheno}

The role of the gravitino in SUSY models is well-known. In global SUSY,
spontaneous breaking produces a massless (spin-1/2)
goldstino with derivate couplings to the SUSY current. This is expressed
in the SUSY generalization of the Goldberger-Treiman formula~\cite{fayet}:
%\bea
%{\cal L}&=&\frac{1}{F} J^{\alpha\mu}\partial_\mu\gravitino_\alpha 
%\nonumber \\
%&=&\frac{i}{4\pi}\bar\lambda^A\gamma^\rho\sigma^{\mu\nu}\partial_\rho
%\gravitino\, F^A_{\mu\nu} \nonumber \\ 
%& &{}+\frac{i\sqrt{2}}{F}\bar\psi_L\gamma^\mu\gamma^\nu\partial_\mu
%\gravitino\, D_\nu\phi + h.c.
%\eea
$$
{\cal L}\sim\frac{1}{F}
\left(\bar\lambda^A\gamma^\rho\sigma^{\mu\nu}\partial_\rho
\gravitino F^A_{\mu\nu} + \bar\psi_L\gamma^\mu
\gamma^\nu\partial_\mu\gravitino D_\nu\phi\right). 
$$
Here $F$ is the $F$-term responsible for the original SUSY-breaking
in the hidden sector; it can be thought of as being the largest
$F$-term in the theory.
The decay width of sparticles into their
spartners plus gravitinos can then be calculated. For example, the width
for a bino, $\bino$, decaying into a photon and gravitino is given
by:
\beq
\Gamma(\bino\to\gamma\gravitino)=\frac{\cos^2\theta_W}{8\pi}\,
\frac{m^5_\bino}{F^2}
\eeq
which becomes larger as $F$ becomes smaller.

Once SUSY is elevated to a local symmetry (\ie, supergravity), the
goldstino is eaten via the super-Higgs mechanism by the massless
(spin-3/2) gravitino. The gravitino acquires longitudinal
components and a mass: $m_{3/2}\simeq F/\mpl$. Since gravity is
so weak, it is only the longitudinal components of $\gravitino$ that
interact. Therefore the results which were derived for the goldstino
hold equally well for the gravitino.

In supergravity models one rarely worries about $\gravitino$. It is too
weak to be produced directly in experiments, and there is no reason why
it should be lighter than any other SUSY partner so that others states
decay into it. (For my purposes here, I am always assuming that there
is a discrete symmetry, like $R$-parity, which makes the lightest
SUSY particle (LSP) absolutely stable.) In gauge-mediated 
models the gravitino is light, roughly $1\ev$ to $1\gev$, making it 
the LSP. It is still too weak to be directly produced in experiments, but
as the LSP, all other SUSY particles must eventually decay into it.
The phenomenology of gauge-mediation, which is otherwise so much like
that of supergravity, has a new component, the search for decays into
gravitinos.

There are two central questions which arise in studying gravitino
phenomenology~\cite{dtw,akkm}. 
First, what is the NLSP (the next-to-lightest SUSY
particle)? Even with light gravitinos, the SUSY states will dominantly decay
via their strong and electroweak interactions, until the
only sparticles left are the NLSPs. The NLSP, having no other route for
its decay, eventually goes to the gravitino plus some other particle(s)
whose identity relies heavily on the type of NLSP present.
The second question is, what is the decay width (or decay length) of the 
NLSP into 
gravitinos? Since the NLSP decay length scales as $F^2$, a measurement
of that length is a direct measurement of SUSY-breaking in the hidden
sector!

The answer to the first question may be model-dependent, but is usually
one of only a few possibilities. For models with only one pair of messenger
fields, and for $\tan\beta$ small to intermediate, the NLSP is a neutralino,
$N_1$, which is itself usually bino-like. For larger multiplicities
of messenger fields, the NLSP(S) are the RH sleptons. But
as $\tan\beta$ increases, the $\stau$ becomes the sole NLSP. For bino
NLSP, we can expect decays most often to $\gravitino+\gamma$; for 
slepton NLSP we can expect decays to the partner lepton, and in 
particular, $\tau$-leptons.

The answer to the second question dictates whether or not the NLSP decays
inside or outside the detector, and if inside, whether the decay length
is long enough to be reconstructed. The various
possibilities are given the following table adapted from
Ref.~\cite{akkm}:
%\begin{table}[t]
%\centering
\begin{tabular}{ccc} & & \\ \hline
NLSP & Decay Length & Signal \\ \hline
~ & Prompt & $\gamma\gamma+\misset$ \\
$N_1\simeq\bino$ & $2^{\rm nd}$ Vertex & $\gamma\gamma+\misset$ \\
~ & Outside & $\misset$  \\ \hline
~ & Prompt & $\ell\ell+\misset$ \\
$\widetilde\ell=\selectron,\smuon,\stau$ & $2^{\rm nd}$ Vertex &
$\ell\ell+\misset$ w/ kinks \\
~ & Outside & Heavy Leptons \\ \hline & & \\
\end{tabular}
%\caption{}
%\label{table:nlsp}
%\end{table}
Here ``prompt'' decays are too close to the vertex to
differentiate, ``$2^{\rm nd}$ vertex'' refers to differentiated
second vertices at which the decay to gravitinos occurred, and ``outside''
means that the decay took place outside of the detector. The signals
are self-explanatory apart from the following note: ``kinks'' are 
sudden turns in the track of a charged particle, in this case occuring
where the invisible $\gravitino$ is being emitted. ``Heavy leptons'' means
that the track will be charged, but not jetty, and will reconstruct to have
a mass far above normal lepton masses. Of course, it is also possible
that more than one option in the table is realized. Either the decay length
could put it on one of the boundaries in the table, or the
lightest slepton and neutralino could be so close in mass that they 
prefer to decay to gravitinos rather than to one another.

\section{Gauge Mediation and Cosmology} \label{sec:cosmo}

In any gauge-mediated model, the gravitino will be the LSP. It mass,
however, is model-dependent. In the Dine-Nelson models we can estimate
the gravitino mass:
\beq
m_{3/2}\simeq\frac{1}{\mpl}\left(\frac{16\pi^2}{g^2_M}\Lambda\right)^2
\simeq20\kev
\eeq
where I have assumed the messenger group coupling, $g_M$, in the last
equality to be $\order(1)$. For direct transmission models, the mass is
larger, about $1\gev$. Finally, though there is no full model at present
that does such, models in which the NLSP decays to gravitinos inside
the detector will have $m_{3/2}\lsim 1\kev$.

Using the standard techniques, the relic abundance of gravitinos present
in the universe today can be calculated.
For gravitinos whose abundances are not diluted by some mechanism, one 
needs~\cite{pp} $m_{3/2}<2h^2\kev$
in order to avoid overclosing the universe. ($h$ is the Hubble
constant in units of 100~km/sec/Mpc.) 
Thus the gravitinos which would be implied by
observing NLSP decay would be cosmologically acceptable and may even be
a useful source of dark matter. 

In the mass range $1\kev\lsim m_{3/2}\lsim 100\kev$, such as
in the Dine-Nelson models, an overabundance of gravitinos results
from late NLSP decay. To wash out this overabundance one would like a period
of late inflation, with the constraint that the reheating temperature, $T_R$,
is less than $\mtwid\sim\mz$ in order to avoid producing more of the NLSP
which would again decay to gravitinos~\cite{mmy}.

In the mass range $100\kev\lsim m_{3/2}\lsim5\gev$, such
as one finds in direct transmission models, the overabundance of
gravitinos is produced through scattering processes: $A+B\to C+\gravitino$.
Once again a period of late inflation can wash out the excess gravitino
abundances, but only as long as $T_R$ does not get
so large as to reproduce the conditions of the gravitino production. In this
case, that means $T_R<10^8\,m_{3/2}$~\cite{mmy}.

Finally for gravitinos with $m_{3/2}\gsim5\gev$, the decay
width $\Gamma(\mbox{NLSP}\to\gravitino)$ is very narrow. The NLSP
doesn't decay until after nucleosynthesis, destroying light nuclei
by photofission. It seems very difficult to avoid
the problems associated with a gravitino in this mass range.

There is one other cosmological concern one might have in gauge-mediated
models which only arises when coupling the model to string theory.
In string theory, there are fields called string moduli whose vevs
parametrize the size of the compactified extra dimensions but whose 
potentials are flat before SUSY is broken. After SUSY-breaking, the
moduli get masses $m\sim m_{3/2}$. 

In the high density, high temperature early universe the potential for
the moduli gets additional contributions proportional to the Hubble constant,
$H$,
and the temperature, $T$, since both break SUSY. But the minimum of the 
moduli potential at finite $H$ and $T$ need not be the same as the minimum
when $H=T=0$. Thus when $H,T$ fall below $m_{3/2}$ the moduli begin falling
towards their true minimum. However, since their couplings to matter are
Planck-suppressed, there is little damping and the moduli begin
oscillating about their minima with amplitudes of $\order(\mpl)$. During
the period of their oscillations, they contribute to the energy density
of the universe through their $(\nabla\phi)^2$ kinetic energy.

If the potential is too shallow and the damping too small, the moduli are
still oscillating today. That energy density would easily overclose the
universe unless some means was found to inflate away the moduli. Such an 
inflation seems difficult to arrange 
for the moduli of the Dine-Nelson models, but
may be possible for the heavier direct transmission moduli~\cite{dgmm}.

\section{Conclusions}

The study of gauge-mediated SUSY-breaking has flowered dramatically over
the last year. Though much
of the early attention on these models focussed on the single
$ee\gamma\gamma+\misset$ event at CDF~\cite{cdf}, the theoretical interest
and justification for these models go well beyond any single experimental
anomaly. Pragmatically, these models provide an alternative measure
for testing experimental sensitivities to SUSY in all its guises. And they
are simple to work with, with only 3 free parameters: $\Lambda$,
$\tan\beta$ and $\log\mmess$. One particularly attractive version of gauge
mediation (the MMM) has only two, $\tan\beta$ being an output.

Idealistically, gauge-mediated models provide an opportunity to do something
which is usually very difficult, and that is to probe
the physics of the hidden sector directly. And by solving the SUSY flavor
problem, they allow interesting flavor physics to occur at scales not far
above the weak scale without inducing large FCNCs. 

Progress in this area
seems to require effort in two directions right now: careful study of the
phenomenology of these models at current and future experiment, with
special attention paid to going beyond the minimal models; and continued
attempts to build realistic models of gauge-mediation which work around
the problems I have outlined here. If progress continues at the same rate this
coming year as it has this past year, a review talk on gauge-mediation at
SUSY-98 would be both very difficult {\sl and} very exciting to give.

\end{document}